# Comparative high-pressure study on rare-earth entropy fluorite-type oxides

Pablo Botella[a,*], David Vie[b] Leda Kolarek[c], Neha Bura[a], Peijie Zhang[a], Anna Herlihy[d], Dominik Daisenberger[d], Catalin Popescu[f], and Daniel Errandonea[a]

[a]*Departamento de Física Aplicada-ICMUV, MALTA Consolider Team, Universitat de Valencia, 46100 Valencia, Spain*
[b]*Institut de Ciència dels Materials de la Universitat de València, Apartado de Correos 2085, E-46071 València, Spain*
[c]*Faculty of Chemical Engineering and Technology, University of Zagreb, 10000 Zagreb, Croatia*
[d]*Diamond Light Source Ltd., Harwell Science & Innovation Campus, Diamond House, Didcot, OX11 0DE, UK*
[f]*CELLS-ALBA Synchrotron Light Facility, Cerdanyola del Vallés, 08290 Barcelona, Spain*

**Keywords:** Entropy varying, X-ray diffraction, Raman spectroscopy, High-pressure

*Corresponding author:
E-mail address: pablo.botella-vives@uv.es

**Abstract**

We report a comparative high-pressure study of two fluorite-type rare-earth oxides with increasing configurational entropy, $(CePr)O_{2-\delta}$ and $(CePrLa)O_{2-\delta}$. Synchrotron-based powder X-ray diffraction and Raman spectroscopy were carried out up to 30 GPa and 20 GPa, respectively. Both compounds retain the cubic fluorite structure throughout the pressure range explored, although an anomaly is observed between 9-16 GPa, characterized by a compressibility plateau and changes in vibrational modes. This behavior is attributed to local lattice distortions and a progressive bond angle bending rather than abrupt phase transitions. In $(CePrLa)O_{2-\delta}$, the onset of amorphization is observed above 22 GPa, highlighting its reduced structural stability. The bulk modulus of both systems shows a slight decrease after the onset of the anomaly, suggesting subtle lattice softening. Raman spectroscopy reveals suppression of the $F_2g$ mode intensity with increasing cationic disorder, and under compression, partial reordering is evidenced by an increase in the RE–O mode intensity. Our results highlight the complex interplay between configurational entropy, cation size, and pressure in determining the structural stability and vibrational properties of rare-earth high-entropy oxides and provide insight into the mechanisms governing their resilience and local disorder under extreme conditions.

## 1. Introduction

Since their first synthesis in 2015 by Rost et al. [1], high-entropy oxides (HEOs) have emerged as a novel class of materials with exceptional structural and functional versatility. These materials, defined by the presence of multiple cationic species in near-equimolar ratios, exhibit remarkable properties such as enhanced thermal stability, mechanical hardness, oxidation resistance, and tunable electrical and magnetic behavior. Rare-earth-based (RE) HEOs [2], in particular, are of growing interest due to the rich electronic configurations of *f*-elements, enabling complex interactions and functionalities relevant to energy, electronics, catalysis, and aerospace applications [3]-[6].

The structural stabilization of these multi-component oxides is governed primarily by configurational entropy, which can lower the Gibbs free energy ($\Delta G = \Delta H - T\Delta S$), thereby promoting the formation of single-phase solid solutions that are otherwise immiscible. This entropic contribution is maximized when the constituent elements randomly and in equimolar proportions occupy equivalent lattice positions, as in the fluorite ($CaF_2$-type) structure, and becomes increasingly dominant with the number of components (N). However, the high degree of chemical disorder can also introduce significant local distortions and raise the internal energy (U), making the balance between entropy and enthalpy a delicate and composition-dependent issue [7],[8].

External pressure provides an additional thermodynamic variable to probe the stability of HEOs. Through the PV term in the enthalpy ($H = U + PV$), pressure can significantly alter the Gibbs energy landscape, stabilizing or destabilizing phases, modifying local environments, and inducing phase transitions. It can also affect oxygen sublattices, induce amorphization, or even change the valence state of the cations [7]-[10]. Hence, high-pressure (HP) studies are crucial to understand the complex interplay between entropy-driven stability and lattice energetics in these systems.

In this work, we present a HP synchrotron XRD and Raman spectroscopy study of $(CePr)O_{2-\delta}$ and $(CePrLa)O_{2-\delta}$, two fluorite-type rare-earth oxides that represent early steps in a compositional series of increasing configurational entropy. By compressing these compounds up to 30 GPa, we aim to investigate how entropy and cation chemistry influence structural response to pressure, including potential phase transitions or amorphization. These compositions were chosen as intermediate references between $CeO_2$ and the high-entropy oxide $(CePrLaYSm)O_{2-\delta}$, which has been already studied under pressure, enabling a systematic analysis of the roles played by entropy, oxidation state, and cation size [7][11]-[17]. Ce and Pr, both capable of adopting a $4^+$ oxidation state under oxidizing conditions, contribute to fluorite phase stabilization, while the introduction of $La^{3+}$ increases entropy and ionic size mismatch. This approach allows us to study the factors governing phase stability and pressure response in complex rare-earth oxide systems.

## 2. Materials and methods

### a. Synthesis of materials

Materials used as reagents in the current investigation were $Ce(CH_3CO_2)_3{\cdot}3H_2O$, $Pr(CH_3CO_2)_3{\cdot}3H_2O$ and $La(CH_3CO_2)_3{\cdot}3H_2O$, analytically pure (99.9%) and used as received from Thermo Scientific Chemicals. The starting Ce-, Pr-, or La-containing solutions were prepared by dissolving their respective salts in distilled water. Then, they were combined to obtain Ce-Pr or Ce-Pr-La source solutions having a total cationic concentration of 0.2 M and a total volume of 25 mL. A small amount of acetic acid was added to the solution after the mixture to ensure long-term stability of the solutions. Droplets of these solutions were flash frozen by projection onto liquid nitrogen and then freeze-dried at a pressure of 1-10 Pa and at a temperature of 228 K in a Telstar Cryodos freeze-dryer. In this way, dried solid precursors were obtained as amorphous loose powders. The final oxides were synthesized by thermal decomposition of the amorphous precursor solids. A sample of the selected precursor (ca. 0.2 g) was placed into an alumina boat and introduced into the furnace. The precursor powder was heated under ambient atmosphere at 5 K $min^{-1}$ to a final temperature of 973 K and was held for a period of 1 h. Then, the solid was cooled, leaving the sample inside the furnace.

### b. EDAX measurements

Rare earth ratios in the solids were determined by energy-dispersive analysis of X-ray (EDAX) on a Thermo Scientific SCIOS 2 scanning electron microscope using an Oxford Ultim Max 170 detector. The operating voltage was 20 kV, and the energy range of the analysis was 0-10 keV.

### c. High-pressure XRD experiments

Angle-dispersive X-ray diffraction (ADXRD) experiments were conducted on the $(CePr)O_{2-\delta}$ and $(CePrLa)O_{2-\delta}$ compounds at the MSPD-BL04 beamline of the ALBA synchrotron. A monochromatic X-ray beam with a wavelength of 0.4246 Å was focused to a 15 μm × 15 μm spot size (full width at half maximum) using Kirkpatrick–Baez mirrors. The diffraction patterns were recorded using a Rayonix SX165 CCD detector placed at a fixed sample-to-detector distance of 260 mm [18].

High-pressure conditions were achieved using a membrane-type Almax diamond anvil cell (DAC) equipped with 500 μm culet diamonds. The pressure chamber was prepared by pre-indenting a stainless-steel gasket and drilling a central hole of approximately 200 μm in diameter and 45 μm in height using an electrical discharge machine. Copper (Cu) powder was loaded alongside the samples and used as an internal pressure calibrant [19]. A 4:1 methanol–ethanol (ME) mixture was employed as the pressure-transmitting medium (PTM) to ensure quasi-hydrostatic conditions throughout the compression cycle [20].

Each sample was compressed in steps of approximately 0.4 GPa up to a maximum pressure of 30 GPa, with diffraction patterns collected at each step. Measurements were also performed during decompression to evaluate reversibility and possible hysteresis effects. The collected 2D diffraction images were azimuthally integrated using the Dioptas software package [21] to obtain conventional one-dimensional diffractograms. Structural analysis of the integrated data was performed using the GSAS-II suite [22], while structural visualizations were generated using VESTA software [23].

### a. High-pressure Raman experiments

Raman measurements were performed using a Horiba LabRam HR800 spectrometer equipped with 1200 and 2400 lines/mm gratings located at Diamond light source, at beamline I15. In this study, the 1200 grating was used in combination with a 20× objective. The system includes confocal spatial filtering, and with an entrance slit set to 100 μm, this configuration results in an effective sampling spot of approximately 5 μm. Excitation was provided by a 523 nm (green) laser with a nominal maximum output power of 500 mW. The power was set to 80% of total power. The total time for each point measurement was 10 min. Precautions were taken to minimize possible laser-induced heating effects during Raman measurements. High-pressure conditions were achieved using a LeToullec-style membrane-driven diamond anvil cell (DAC) with a 4-pin configuration. The anvils employed 400 μm culets and were of Boehler-Almax type, made from synthetic type IIa diamonds. The anvils were mounted on tungsten carbide (WC) seats. A ruby chips-sphere along with the sample was introduced to monitor the pressure [24]. The PTM used was 4:1 ME for a better comparison between XRD and Raman measurements.

## 3. Results and discussion

### a. Ambient conditions structural characterization

The structural integrity and phase purity of the synthesized fluorite-type compounds, $(CePr)O_{2-\delta}$ and $(CePrLa)O_{2-\delta}$, were first evaluated through Rietveld refinement of their powder XRD patterns, as shown in **Figure 1**. The good agreement between the experimental data and the calculated profiles confirms the successful formation of single-phase fluorite structures, assigned to the cubic space group $Fm\bar{3}m$ (No. 225), in both compositions [2]. The sharp, symmetric Bragg reflections and minimal residuals in the difference curves further support the high crystallinity and structural homogeneity of the samples.

Crystallite sizes were estimated using the Scherrer equation, considering up to eight distinct reflections in each diffraction pattern. The calculated average crystallite sizes were $18.3 \pm 0.2$ nm for $(CePr)O_{2-\delta}$ and $11.3 \pm 0.2$ nm for $(CePrLa)O_{2-\delta}$, indicating nanoscale grain dimensions consistent with the synthesis approach.

To verify the homogeneity and nominal composition of the as-synthesized $(CePr)O_{2-\delta}$ and $(CePrLa)O_{2-\delta}$ powders, EDAX measurements were performed. For each sample, five independent measurements were conducted at different regions of the powder to account for possible local compositional variations. The average atomic percentages obtained from these measurements are summarized in **Table 1**. The results confirm that the elemental ratios are consistent with the intended stoichiometry (equimolar distribution) to maximize the entropy level, supporting the successful synthesis of homogeneous multicomponent rare-earth oxide phases. Furthermore, the configurational mixing entropy was estimated **Table 1**, and presented in **Figure 2a**, together with the values for $CeO_2$ and $(CePrLaYSm)O_{2-\delta}$, illustrating the low- and medium-entropy character of the studied compounds. The configurational mixing entropy ($\Delta S_{mix}$) was calculated using the standard expression for ideal mixing of N cationic species:

$$\Delta Smix = -R\sum_{i=1}^{N} x_i\, lnx_i$$

where R is the universal gas constant and $x_i$ is the molar fraction of each cationic species. For equimolar binary $(CePr)O_{2-\delta}$, this yields $\Delta S_{mix}$ = 0.69R, and for ternary $(CePrLa)O_{2-\delta}$, $\Delta S_{mix}$ = 1.09R. We have used the ideal values for the molar fraction 0.5 for $(CePr)O_{2-\delta}$, and 0.33 for $(CePrLa)O_{2-\delta}$ supported by the measured values by EDAX which are quite close to those.

To further explore the structural implications of increasing configurational entropy, **Figure 2b** presents a comparative analysis of the lattice parameters and the effective crystal radius of the rare-earth cationic sublattice. The dataset includes reference values for binary $CeO_2$, as well as for the higher-entropy compound $(CePrLaYSm)O_{2-\delta}$, thereby spanning a wide entropy gradient from low to very high configurational complexity.

The observed trend reveals a systematic expansion of the fluorite lattice due to the progressive substitution of smaller tetravalent cations (e.g., $Ce^{4+}$) with larger trivalent ions such as $Pr^{3+}$, $La^{3+}$, $Y^{3+}$, and $Sm^{3+}$. This behavior aligns with expectations based on the Shannon crystal radii for 8-fold coordination, confirming the sensitivity of the fluorite-type structure to variations in cationic size and composition. In particular, the incorporation of $La^{3+}$ leads to a lattice expansion from 5.416(3) Å ($CeO_2$) to 5.523(5) Å in $(CePrLa)O_{2-\delta}$. For $(CePr)O_{2-\delta}$, the lattice parameters suggest the possible presence of $Pr^{4+}$ ions, despite the precursor adopting the hexagonal $Pr_2O_3$ structure. A mixed-valence state of $Pr^{3+}/Pr^{4+}$ has previously been reported during the synthesis of $(CePrLaYSm)O_{2-\delta}$ [2], and is believed to contribute, alongside trivalent substitutions, to the formation of oxygen vacancies that help maintain charge neutrality in the system. Although direct spectroscopic techniques (e.g., XPS or XANES) were not employed in this study, the presence of $Pr^{4+}$ is inferred based on charge balance considerations and by analogy with similar fluorite-type rare-earth oxides [2,7]. In these systems, oxygen vacancies and mixed-valence states often coexist, with redox-active elements like Pr facilitating local charge compensation. Under compression, the observed enhancement of the RE–O Raman mode and the concurrent evolution of the VO band (see further in the text) may reflect pressure-induced redistribution of vacancies and subtle redox changes. Nevertheless, these hypotheses require confirmation by future studies involving direct spectroscopic or computational approaches.

**Figure 2c** presents a structural visualization of the $(CePrLa)O_{2-\delta}$ compound, illustrating its face-centered cubic structure, which belongs to the $Fm\bar{3}m$ space group (No. 225). The $CaF_2$-type structure of the rare earth oxides consists of $REO_8$ hexahedra (cubes) with a cubic close-packed arrangement, in which each RE atom is surrounded by eight O atoms forming eight RE–O bonds. All the $REO_8$ cubes connect with each other by sharing one edge [26]. The model highlights the random distribution of Ce, Pr, and La atoms within the cationic sublattice, a hallmark of configurational disorder. Oxygen atoms, depicted in red, occupy the fluorite-type coordination sites at the Wyckoff position (¼, ¼, ¼), while the rare-earth cations share the Wyckoff position (0, 0, 0), reflecting a mixed occupancy that underpins the entropic stabilization. This structural model evidences the formation of a homogeneous solid solution, where configurational entropy plays a crucial role in maintaining phase stability despite the ionic size mismatch and charge differences among the constituent cations.

The differences observed in the lattice parameters of the studied compounds at low- and intermediate-entropy compositions further suggest the presence of local lattice distortions, strain fields, and potentially defect formation resulting from increased chemical disorder. Such deviations are typical of high-entropy oxide systems, where the complex cationic environment disrupts ideal lattice packing and generates a distribution of local bonding environments [1][2]. These findings emphasize that entropy-driven stabilization is intrinsically linked to atomic-scale structural frustration, which may have important implications under high-pressure conditions.

## b. Structural characterization under high-pressure conditions

Selected XRD patterns of $(CePr)O_{2-\delta}$ and $(CePrLa)O_{2-\delta}$ measured under high-pressure (room-temperature) conditions are shown in **Figure 3(a) and (b)**. The maximum pressure reached in the experiments was about 30 GPa. Qualitatively, it can be observed that both samples retain the cubic fluorite structure throughout the entire pressure range explored.

Notably, the pressure has a more pronounced effect on $(CePrLa)O_{2-\delta}$, due to the higher lattice distortion introduced by the incorporation of $La^{3+}$ cations. This is evidenced by the broader peaks and reduced reflection intensities when compared to $(CePr)O_{2-\delta}$ (note that the reflection (111) of all patterns is normalized to 1 and vertically offset for better comparison). Three distinct regions can be identified based on peak evolution with pressure, with region II showing the least variation, as will be further discussed in the text. The recovered patterns also exhibit the fluorite structure with similar features to the starting materials, highlighting the high structural stability of these compounds, likely due to entropy-driven effects.

The studied compositions are dominated by $CeO_2$, which plays a key role in stabilizing the single-phase in high-entropy oxide structure and provides a $CaF_2$-type host lattice for the other rare-earth cations [2]. Therefore, it is reasonable to expect that their high-pressure behavior will resemble that of pure $CeO_2$ [7]. Nevertheless, no signs of the structural phase transition typically reported for $CeO_2$ around 30–35 GPa were detected in our experiments [11]-[17]. This absence can be explained by considering the limitation of the pressure range studied, the nature of the PTM and the intrinsic configurational entropy of our low- and medium-entropy oxide systems, assuming comparable particle sizes to those in previous studies. In our case, the use of a 4:1 ME mixture as PTM, which is known to become non-hydrostatic above ~12 GPa [20], might have promoted the onset of a phase transition if the structure were inherently unstable, as non-hydrostatic stress conditions tend to reduce the transition pressure [27].

However, the continued stability of the fluorite structure in $(CePr)O_{2-\delta}$ and $(CePrLa)O_{2-\delta}$ suggests that the configurational entropy introduced by cationic disorder enhances structural resilience against pressure-induced transitions. By contrast, in the higher-entropy oxide $(CePrLaYSm)O_{2-\delta}$ [7], a loss of long-range order (amorphization) is observed around 16 GPa, indicating that beyond a certain level of chemical complexity, the stabilizing entropy contribution becomes less effective due to excessive disorder. Based on this, one might expect our samples, being of lower entropy, to undergo a transition at a pressure between ~16 GPa and the 30–35 GPa typical of pure $CeO_2$ [11][17]. However, the observed behavior suggests a shift to even higher pressures, or at least stability comparable to $CeO_2$.

The absence of a clear phase transition up to 30 GPa reflects a balance between entropy stabilization and the intrinsic resistance of the fluorite framework to deformation. In the case of $(CePrLa)O_{2-\delta}$, however, a broad background contribution becomes apparent above ~22 GPa, particularly around the first two Bragg reflections (see **Figure 4**), indicating the onset of partial amorphization. This feature grows with pressure and suggests a progressive collapse of intermediate-range order within the fluorite lattice. Upon decompression, the broad contribution disappears, confirming that the amorphization is reversible. This behavior points to a pressure-driven structural frustration that does not involve irreversible breakdown, but rather a dynamic reorganization of the lattice under stress. Contrary to the observation in $(CePrLaYSm)O_{2-\delta}$, where amorphization was complete at 16 GPa and irreversible at ambient conditions [7], the amorphization observed in $(CePrLa)O_{2-\delta}$ is only partial and fully reversible upon decompression.

It is worth noting that pushing the experiments beyond this pressure range was considered too risky for the instrumentation employed in this study. Before concluding this part of the discussion we would like to note that the stability of $(CePr)O_{2-\delta}$ and $(CePrLa)O_{2-\delta}$ contrast with behavior of ordered $CeO_2$ based alloys like $(CeZr)O_2$ which adopts a cubic pyrochlore structure and undergoes a phase transition at 5 GPa [28]. Pyrochlore can be described as an ordered derivative of the fluorite structure. In pyrochlore $(CeZr)O_2$, in contrast with fluorite oxides, pressure promotes the formation of oxygen vacancies which favors the observed transition [29].

The pressure-dependent structural evolution of the synthesized compounds was investigated through volume–pressure (V–P) measurements, as shown in **Figure 3 (c)**. The compression behavior of $(CePr)O_{2-\delta}$ and $(CePrLa)O_{2-\delta}$ is compared with that of related rare-earth oxides reported in the literature [7][12]-[14][16]. Both synthesized compounds exhibit a smooth and continuous volume reduction up to approximately 30 GPa (region I and III on **Figure 3 (c)**), indicating the absence of any abrupt structural phase transitions within this pressure range. However, a noticeable deviation from linear compression behavior is observed between 9 and 16 GPa (region II in **Figure 3(c)**), where a plateau and non-uniform volume decrease suggests the presence of local lattice distortions [7]. This anomaly may correspond to a subtle change in the first derivative of the volume–pressure curve, potentially indicative of a second-order isostructural transition or a gradual evolution in compressibility. Nevertheless, the preservation of the fluorite-type cubic symmetry, the absence of symmetry breaking or new Bragg reflections, and the reversibility observed upon decompression all support an interpretation based on continuous internal rearrangements rather than a distinct thermodynamic phase transition. The green-shaded region in the plot marks the pressure range (~11–13 GPa) associated with the glass transition of the 4:1 ME PTM [20]. While deviations from ideal hydrostatic conditions in this range may influence compressibility, the onset of the anomaly at lower pressures (~9 GPa) suggests that it is primarily intrinsic and not triggered by the non-hydrostatic behavior of the PTM beyond 12 GPa.

Notably, the compressibility of $(CePrLa)O_{2-\delta}$ is slightly higher than that of $(CePr)O_{2-\delta}$, consistent with its larger initial unit cell volume and the presence of the larger $La^{3+}$ cation. This trend aligns with the expected behavior based on ionic size and supports the notion that increasing configurational entropy through cation substitution can subtly modulate the mechanical response of the fluorite lattice.

Comparison with literature data for $CeO_2$ [16], $PrO_2$ [16], and bulk $CeO_2$ [14] reveals that the synthesized $(CePr)O_{2-\delta}$ compound follows a similar compression trend to both $CeO_2$ and $PrO_2$. However, unlike the findings reported by Gerward et al. [16], no anomalies were observed in their study. The anomalies detected in our samples appear to be closely related to the crystalline size of the particles, particularly when they are in the nanometer range, as also noted in the works of Wang [12] and Rainwater [13]. These studies report behaviors that resemble our observations, including a slight volume expansion under pressure in Sm-doped $CeO_2$ and pure $CeO_2$ when silicon oil is used as PTM. However, in those cases, the anomalies typically emerge after the PTM undergoes a glass transition, which differs from our findings.

Rainwater et al. [13], partially attributed these effects to the high non-hydrostaticity of the silicone oil used as PTM. In contrast, Wang et al. [12], proposed a dual-structure model to explain the anomalous volume expansion observed in nanocrystalline $CeO_2$ under pressure. According to their model, the material consists of a rigid amorphous shell surrounding a relatively soft crystalline core. This configuration can lead to an apparent negative compressibility when subjected to non-hydrostatic conditions, particularly with silicone oil as the PTM. Notably, such behavior was not observed when more hydrostatic PTMs, such as ME mixtures, were used. However, a pressure-induced plateau is observed even when ME is employed as the PTM, including in the high-entropy compound $(CePrLaYSm)O_{2-\delta}$ [7], which exhibits a plateau beginning at approximately 9 GPa and transitions into an amorphous phase beyond 16 GPa. According to Cheng et al. [7], this pressure-induced plateau arises from a progressive structural disorder that begins around 9 GPa, as revealed by in situ synchrotron X-ray diffraction studies. Instead of further compressing the lattice through bond shortening, the structure accommodates pressure mainly by bending the interatomic bond angles within and between the $REO_8$ polyhedral units. While the rigid RE–O bonds remain nearly constant, the continuous distortion of bond angles reduces the long-range crystallographic coherence. As a result, the diffraction peaks cease shifting and gradually broaden and weaken, indicating the onset of topological disorder. This process culminates in the collapse of the long-range order and the formation of an amorphous phase above ~16 GPa. In addition, the studied compound $(CePrLa)O_{2-\delta}$ shows a broader pressure range affected by structural anomalies, spanning from 9 to 16 GPa, as illustrated in **Figure 3(c), region II**. This compound exhibits a larger unit cell volume compared to $(CePrLaYSm)O_{2-\delta}$, which can be attributed to the higher concentration of $La^{3+}$ cations incorporated into the lattice. These findings underscore the critical role of both particle size and PTM characteristics in determining the high-pressure behavior of rare-earth oxides like $(CePr)O_{2-\delta}$ and $(CePrLa)O_{2-\delta}$. Nevertheless, further complementary measurements are necessary to fully elucidate the mechanisms behind these anomalies.

To determine the bulk modulus of the studied compounds, the pressure–volume (P–V) data were fitted using a second-order Birch–Murnaghan (BM) equation of state (EoS). Data points corresponding to regions with structural anomalies were excluded from the fitting. As a result, two sets of bulk modulus values were obtained for each compound: one for the low-pressure region (region I) and one after the structural anomaly.

The bulk moduli for region I (pressure points from 0 to 6 GPa) are shown in **Figure 5**, alongside those of the precursor compounds for comparison [12][14][16][30]. The refined EoS parameters for $(CePr)O_{2-\delta}$ are: $V_0$ = 157.87(2) Å$^3$, $B_0$ = 205.1(19) GPa, with fixed $B_0$' = 4, and a calculated second derivative $B_0$'' = -0.019 GPa$^{-1}$. For $(CePrLa)O_{2-\delta}$, the values are: $V_0$ = 169.68(6) Å$^3$, $B_0$ = 190(4) GPa, $B_0$' = 4, $B_0$'' = -0.0204 GPa$^{-1}$. As expected, the bulk modulus of $(CePr)O_{2-\delta}$ lies between that of $CeO_2$ (its host lattice) and Pr-containing oxides, reflecting the intermediate nature of its composition. In contrast, $(CePrLa)O_{2-\delta}$ exhibits a lower bulk modulus, shifting closer to the values typically found in La-containing oxides, although still within the range of praseodymium oxides. This reduction is attributed to the presence of the additional (third) cation, which distributes the lattice strain among more species and thus reduces the individual influence of each rare-earth element on the overall stiffness of the structure. After the onset of structural anomalies (pressure points from 12 to 30 GPa for $(CePr)O_{2-\delta}$ and from 15 to 30 GPa for $(CePrLa)O_{2-\delta}$), the bulk moduli of $(CePr)O_{2-\delta}$ and $(CePrLa)O_{2-\delta}$ decrease slightly to 182.3(11) GPa and 181(3) GPa, respectively. This suggests that structural modification results in a slight enhancement of the compressibility of the lattice.

### c. High-pressure conditions vibrational characterization

In order to gain deeper insight into the structural characteristics of the studied compounds, Raman spectroscopy was performed at ambient conditions and under high pressure at room temperature. The ambient-condition Raman spectra are shown in **Figure 6**. Two broad features are observed: the first corresponds to the $F_2g$ symmetric vibration mode of the eight-fold coordinated RE–O bond, typically located around ~450 cm$^{-1}$, while the second, at ~600 cm$^{-1}$, is attributed to the oxygen vacancy ($V_O$) band commonly found in high-entropy oxides (HEOs) [7]. In the pure $CeO_2$ system, the $F_2g$ mode is the most intense feature, but its intensity progressively decreases as more cations are introduced into the lattice. This trend is evident when comparing $(CePr)O_{2-\delta}$ and $(CePrLa)O_{2-\delta}$ to the more chemically disordered $(CePrLaYSm)O_{2-\delta}$, where the $F_2g$ peak is almost completely suppressed. The weakening of the $F_2g$ mode is attributed to the increasing degree of local symmetry breaking and severe lattice distortion, resulting from the high chemical disorder inherent to multicomponent systems.

**Figure 7 (a)** shows the representative Raman spectra of $(CePr)O_{2-\delta}$ during compression, showing the evolution of the RE–O vibrational mode (~450 cm$^{-1}$) and the oxygen vacancy ($V_O$) band (~600 cm$^{-1}$). A gradual blue-shift and broadening of both peaks are observed with increasing pressure, along with intensity changes. In the case of $(CePrLa)O_{2-\delta}$ (**Figure 7 (b))**, the Raman spectra show similar trends to those of $(CePr)O_{2-\delta}$, although the intensity of the RE–O mode is generally lower. This reduction is consistent with an increased degree of lattice distortion and higher cationic disorder arising from the presence of $La^{3+}$ in the structure. The $V_O$ band undergoes several notable changes during compression, including variations in intensity and peak shape. As observed in Figure 6(a), two distinct peaks are present at ambient pressure, but they progressively merge into a single broad band under increasing pressure. This behavior suggests that the oxygen vacancies evolve under compression, potentially altering their local environment, distribution, and concentration. This behavior may be associated with the relatively high compressibility of the vacancy sites, as well as pressure-induced changes in the oxidation state of praseodymium, particularly the conversion from $Pr^{3+}$ to $Pr^{4+}$. These factors contribute to the evolving local environment and increased lattice distortion during compression [7].

In **Figure 7 (c)** shows the pressure evolution of the Raman modes under pressure as well as the intensity ratio ($I_{RE\text{-}O}/I_{VO}$). As the spectra exhibit very broad bands, the peak positions were estimated by tracking the intensity maxima to qualitatively follow their evolution under pressure. The $V_O$ band exhibits a linear shift under compression with no notable variation. In contrast, the RE-O band displays a clear change in slope around 10-11 GPa, which correlates with the onset of bond distortion observed in the XRD data. Interestingly, both compounds exhibit an enhanced RE–O mode intensity upon compression, eventually surpassing the intensity of the $V_O$ band. This behavior may indicate a pressure-induced rearrangement or ordering in the local environment, where the RE–O framework becomes more dominant relative to the vacancy-related features. However, upon decompression, these changes are largely reversible, with both the RE–O and $V_O$ modes returning to their original positions and intensities, indicating elastic structural behavior and recovery of the initial local environment.

## 4. <u>Conclusions</u>

In conclusion, the synthesized $(CePr)O_{2\text{-}\delta}$ and $(CePrLa)O_{2\text{-}\delta}$ compounds demonstrate remarkable structural robustness under high pressure, retaining the fluorite-type cubic symmetry up to 30 GPa without undergoing any abrupt phase transitions. Both materials exhibit a distinctive compression anomaly in the 9–16 GPa range, observed as a plateau in their compressibility and a slope change in the RE-O Raman mode. This behavior is attributed to internal structural rearrangements, particularly distortions in bond angles, rather than a conventional crystallographic transition. In $(CePrLa)O_{2\text{-}\delta}$, a reversible pressure-induced amorphization is observed above 22 GPa, as evidenced by the appearance and subsequent disappearance of a broad background signal in the low-angle diffraction region upon decompression. The bulk modulus values correlate with composition, with $(CePr)O_{2\text{-}\delta}$ showing greater stiffness compared to $(CePrLa)O_{2\text{-}\delta}$, while both materials exhibit a slight increase of compressibility beyond the anomaly. Raman spectroscopy further reveals the suppression of the $F_{2g}$ vibrational mode at ambient conditions due to cationic disorder, and its partial recovery under pressure, indicating improved local ordering upon compression. Meanwhile, the VO-related band remains structurally stable but evolves in intensity and shape, likely due to pressure-induced changes in $Pr^{3+}/Pr^{4+}$ redox states and vacancy redistribution. Upon decompression, both structural and vibrational features largely recover, suggesting the changes are elastic and reversible. Overall, these findings highlight the critical influence of configurational entropy and multicationic engineering in governing the high-pressure mechanical and vibrational behavior of rare-earth oxides, paving the way for their design and optimization in extreme environments.

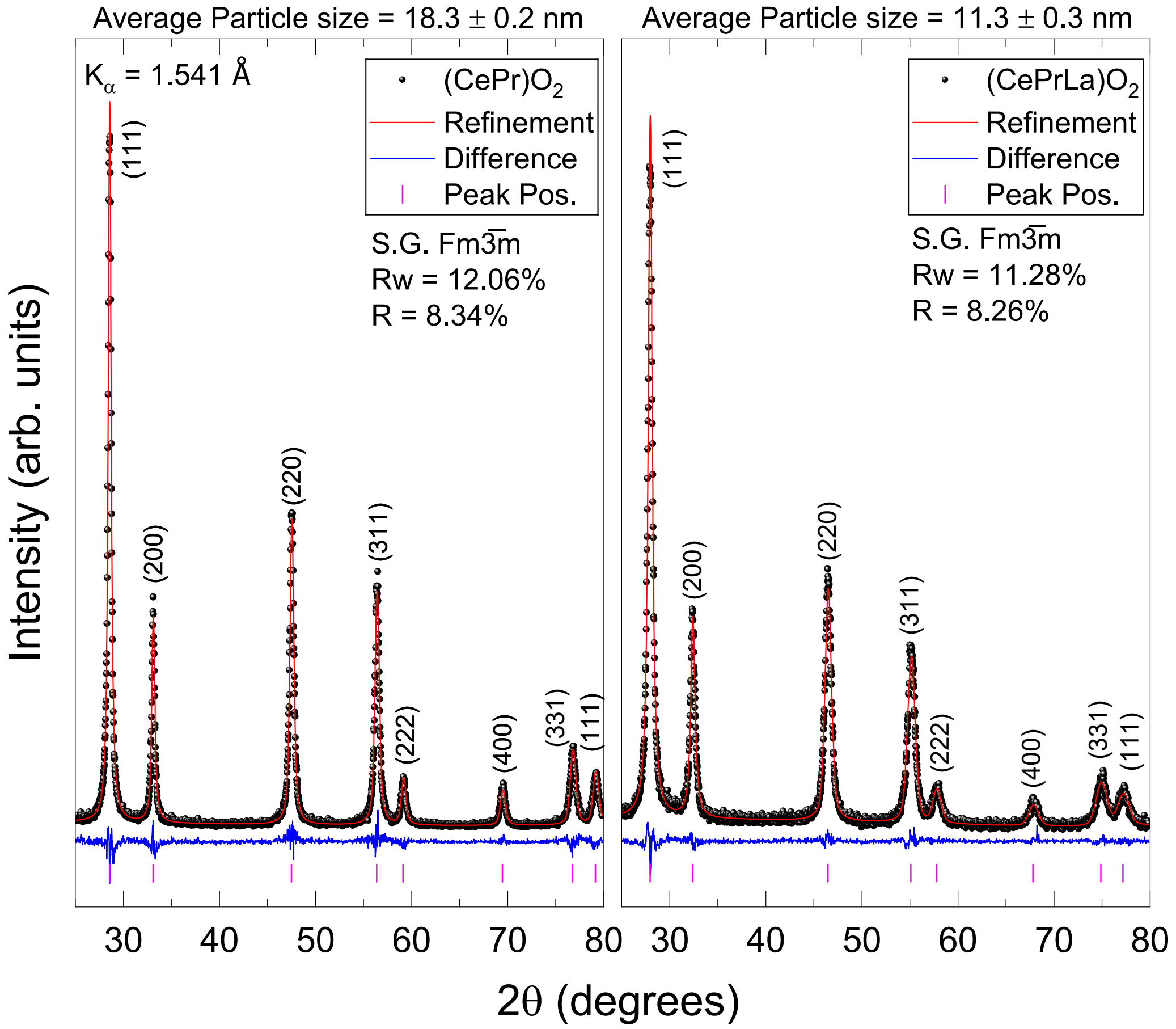

**Figure 1**: Rietveld refinement of the powder XRD pattern of the as-synthesized compounds. The experimental data are shown as black dots, the fitted pattern is indicated by the red line, and the difference curve (experimental minus calculated) is displayed in blue. The vertical purple ticks represent the expected Bragg peak positions for the fluorite-type structure. Average particle size has been estimated by the Scherrer equation.

**Table 1**: EDAX results of the as-synthesized samples. Values represent the average of five measurements at distinct points on the sample surface.

| | **Theoretical** | **Ce** | **Pr** | **La** | **$\Delta S_{mix}$** |
|---|---|---|---|---|---|
| **$(CePr)O_{2-\delta}$** | 50 | 49.5 ± 0.3 | 50.5 ± 0.3 | | 0.69R |
| **$(CePrLa)O_{2-\delta}$** | 33 | 32.1 ± 0.3 | 31.9 ± 0.3 | 35.97 ± 0.3 | 1.09R |

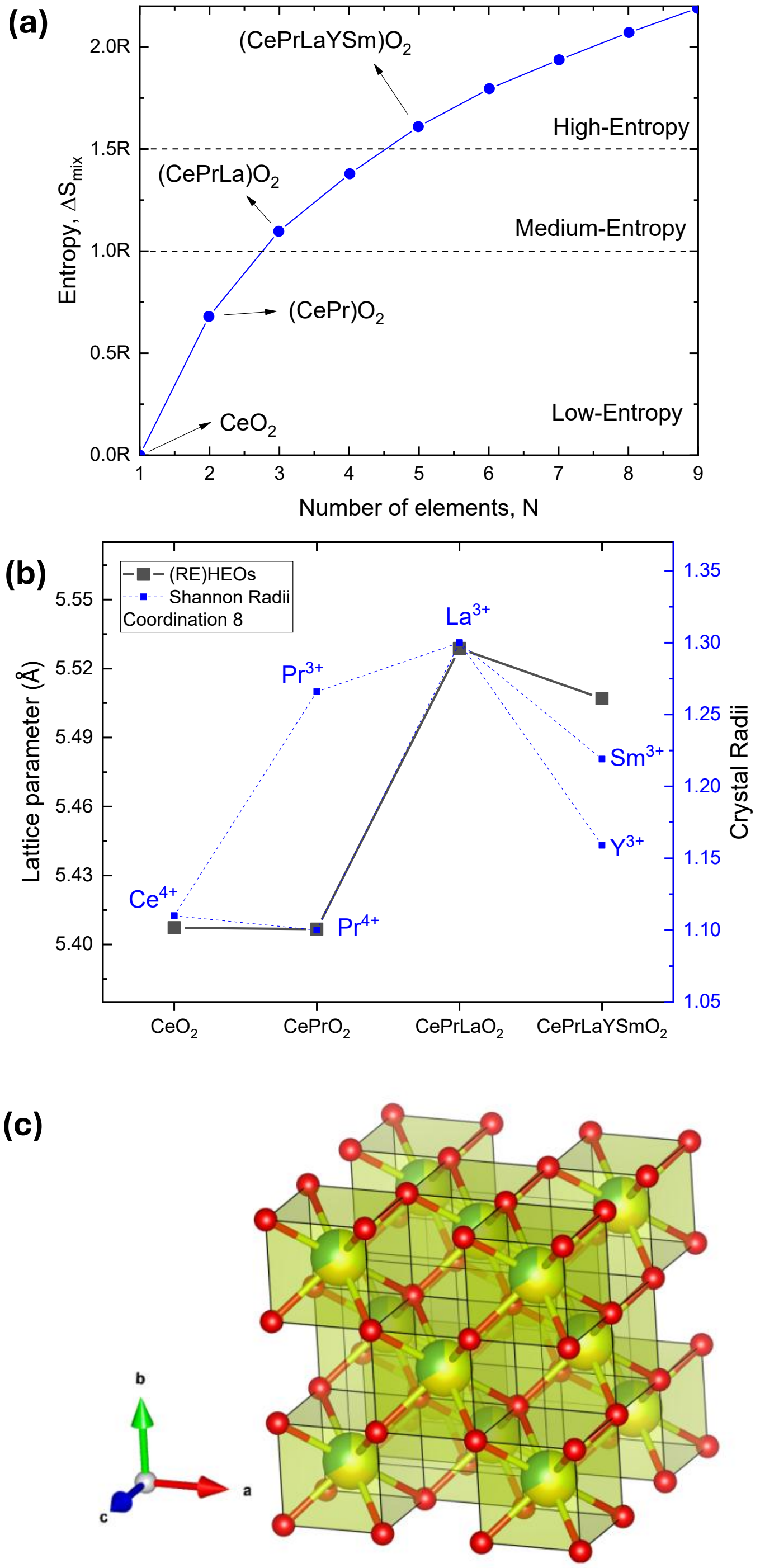

**Figure 2**: (a) The relationship between entropy value, the number of elements and the distribution of low, medium and high entropy ceramics [25]. (b) Comparative lattice parameter versus ionic radii of the constituent elements. Error bars are within the symbol size. (c) Polyhedral crystal structure representation of the (CePrLa)$O_2$. Oxygen atoms are represented by red atoms. Partially colored atoms are equimolarly represented by Ce, Pr or La.

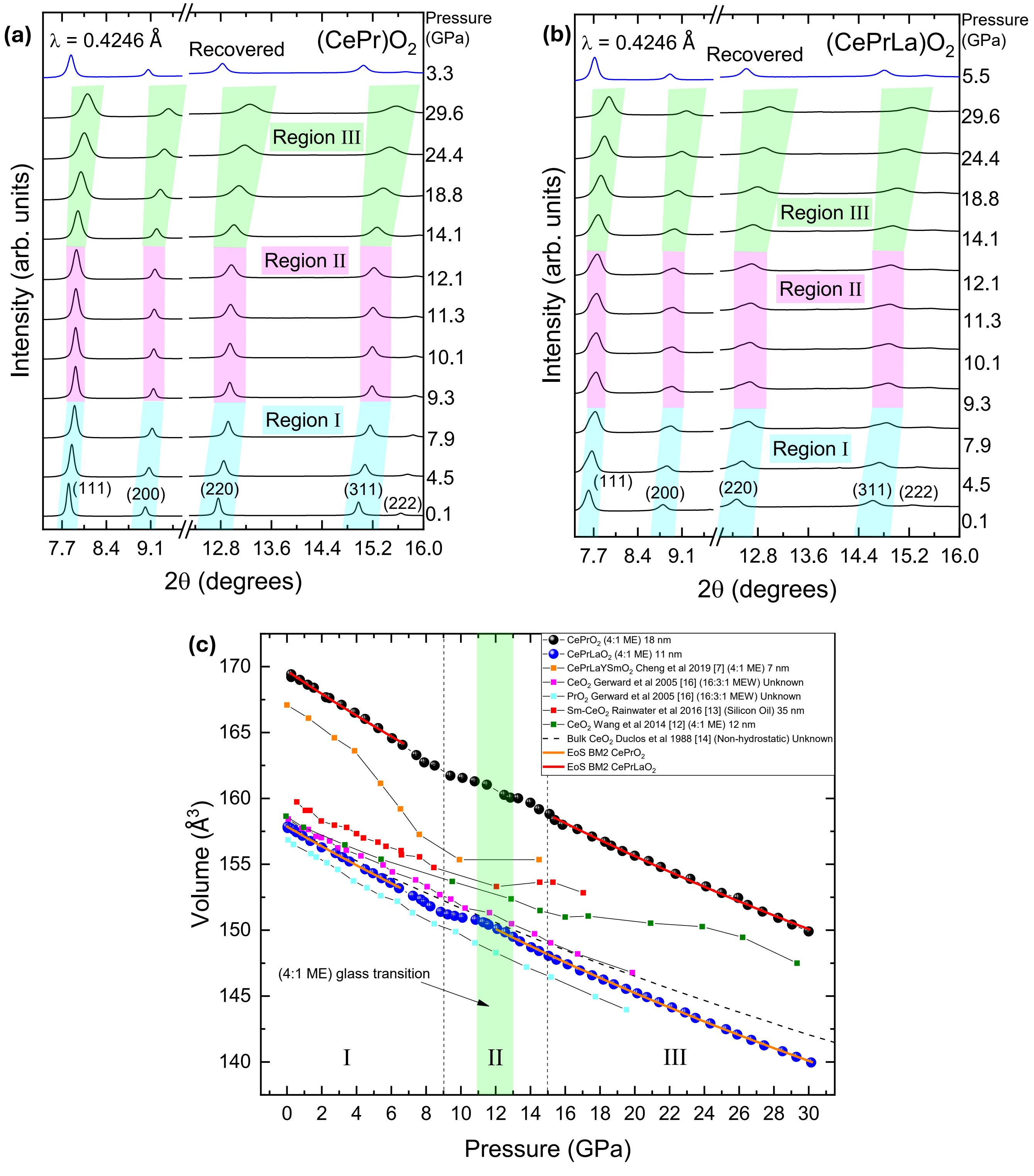

**Figure 3**: (a) Representative X-ray patterns (λ=0.4246 Å) of $CePrO_2$ and (b) of $CePrLaO_2$ at ambient temperature on compression. (c) Unit-cell volume evolution under pressure of both compounds compared to relevant data from literature. The dashed line represents the 2$^{nd}$ order BM EoS fit.

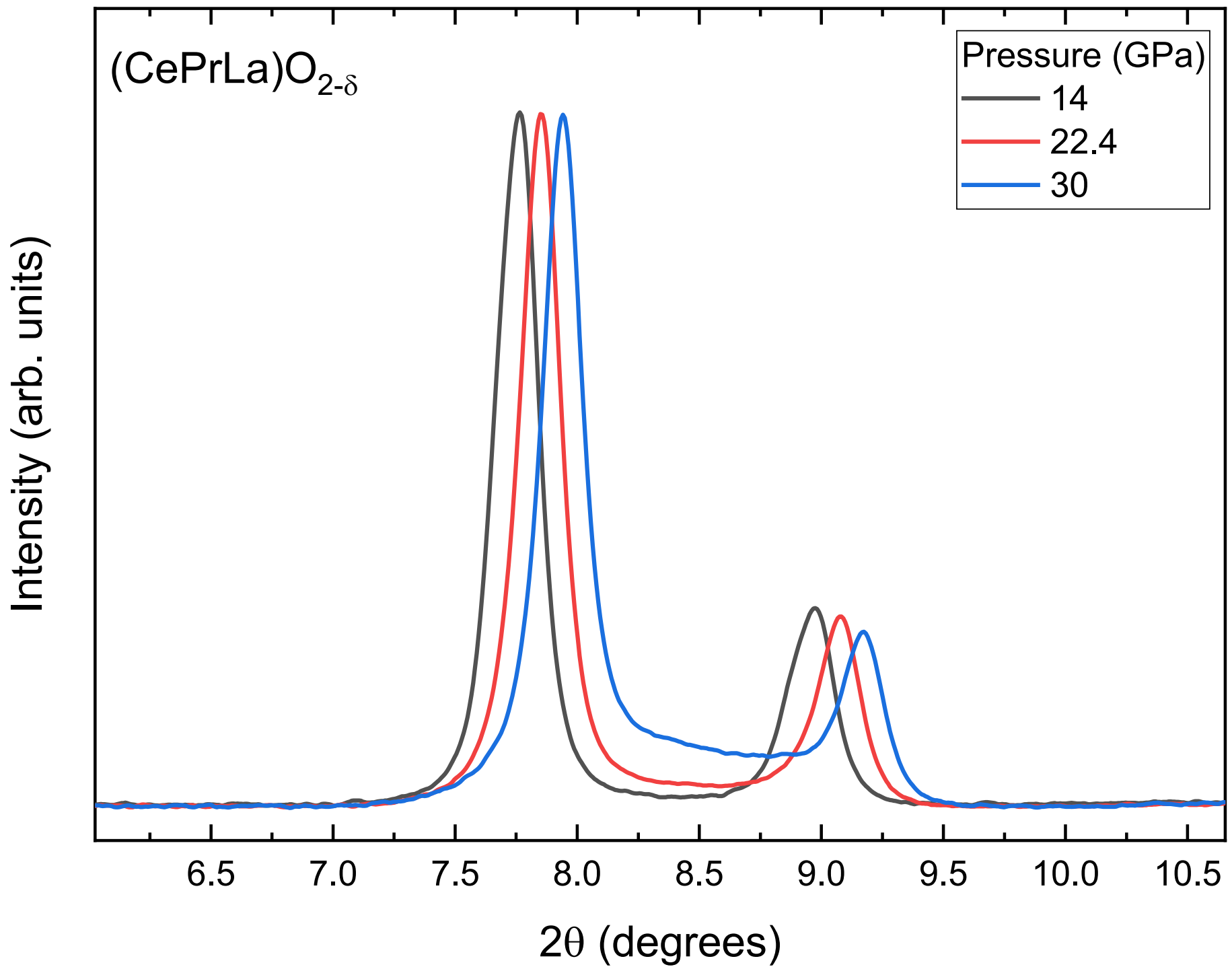

**Figure 4:** XRD patterns at selected pressure showing the partial amorphous contribution.

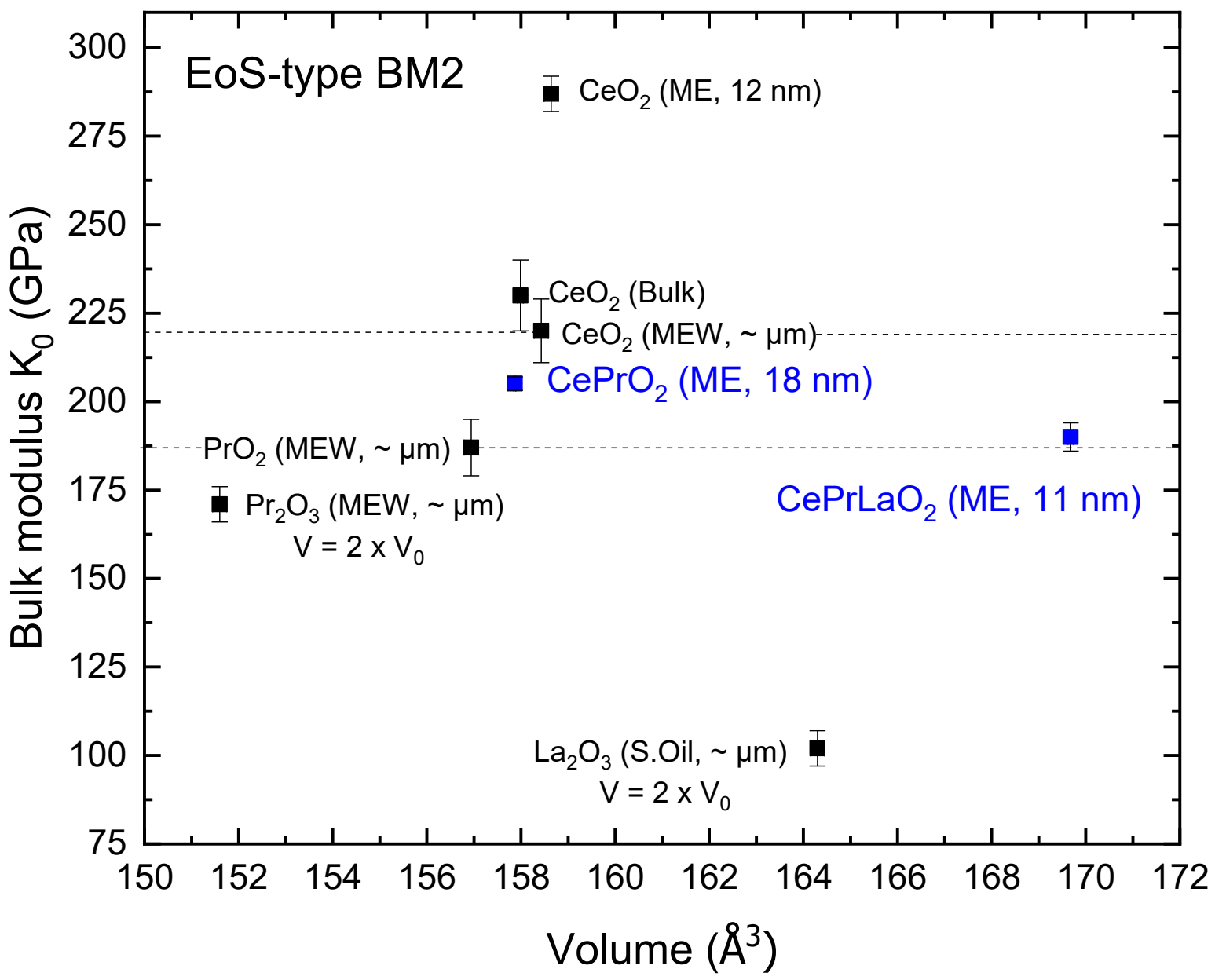

**Figure 5**: The bulk modulus of the $CePrO_2$ and $CePrLaO_2$ compounds using different PTMs as indicated compared to their precursor's bulk modulus [12][14][16][30].

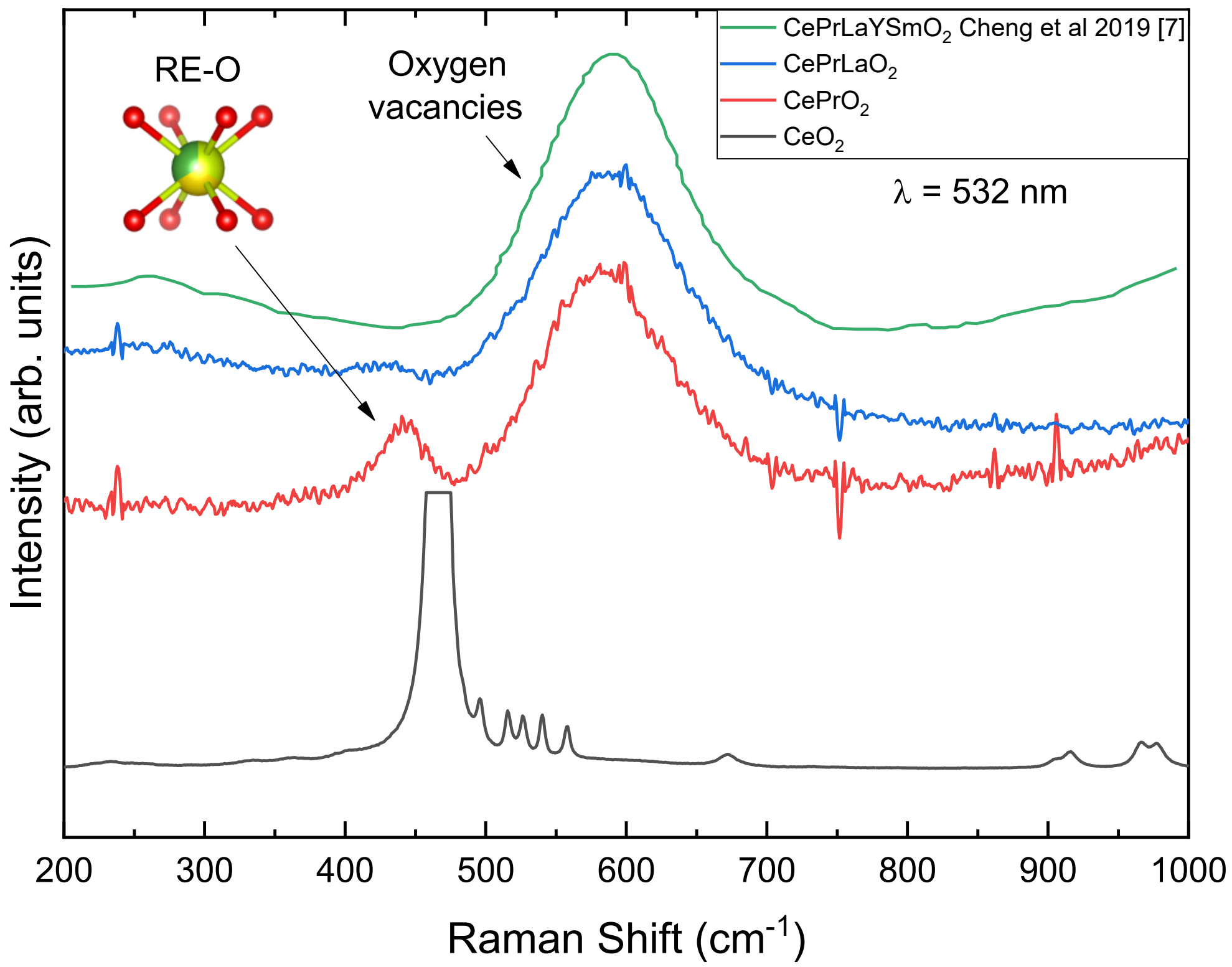

**Figure 6**: Ambient conditions Raman spectra of the $CePrO_2$ and $CePrLaO_2$ along with $CeO_2$ and $CePrLaYSmO_2$ compounds.

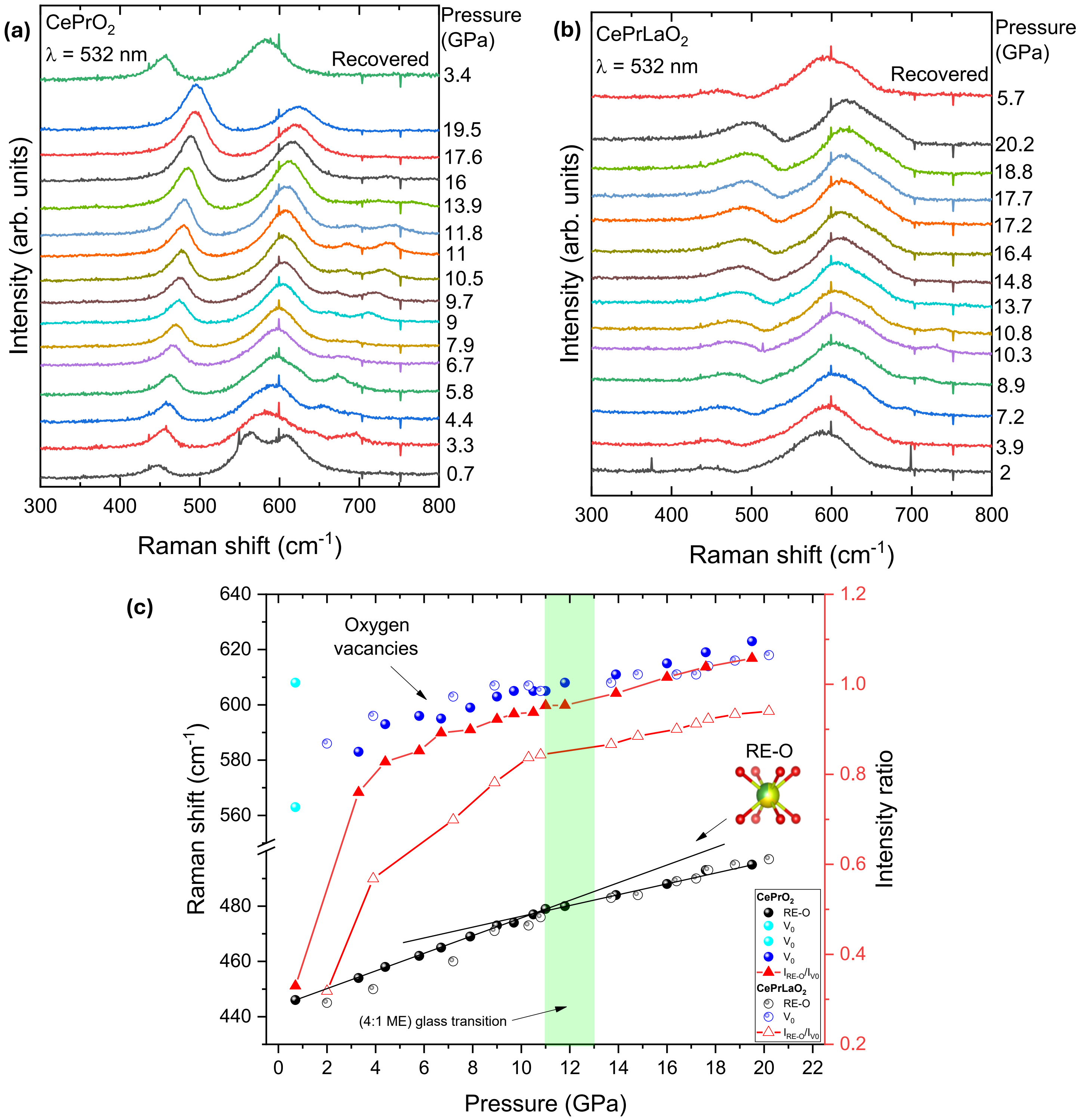

**Figure 7**: (a) Representative Raman spectra (λ=532 nm) of $CePrO_2$ and (b) of $CePrLaO_2$ at ambient temperature on compression. (c) Pressure dependence of the Raman shift for the two main features related to RE-O vibrations and oxygen vacancies. The evolution of their intensity ratio ($I_{RE-O}/I_{V0}$) with pressure is also shown. A straight line has been added to the RE–O mode data to aid visualization of the change in slope.

## Author Contributions

**P. Botella:** Conceptualization, Formal analysis, Methodology, Writing - original draft, Writing - review & editing.
**D. Vie:** Formal analysis, Synthesis, Writing - review & editing.
**L. Kolarek:** Formal analysis, Methodology, Writing - review & editing.
**N. Bura:** Data acquisition, Writing - review & editing.
**P. Zhang:** Data acquisition, Writing - review & editing.
**A. Herlihy:** Data acquisition, Writing - review & editing.
**D. Daisenberger:** Data acquisition, Writing - review & editing.
**C. Popescu:** Data acquisition, Writing - review & editing.
**D. Errandonea:** Formal analysis, Writing - original draft, Writing - review & editing.

## Conflicts of interest

The authors declare that they have no known competing financial interests or personal relationships that could have appeared to influence the work reported in this paper.

## Acknowledgments

The authors acknowledge financial support from the Spanish Research Agency (AEI) and Spanish Ministry of Science, Investigation y Universidades (MCIU) under grant PID2022-138076NB-C41 (DOI: 10.13039/501100011033). This work was also supported by Generalitat Valenciana (GVA) under grants CIPROM/2021/075 and MFA/2022/007. This study forms part of the Advanced Materials program and is supported by MCIU with funding from European Union Next Generation EU (PRTR-C17.I1) and by GVA. P.B. and D.E. thank GVA for the Postdoctoral Fellowship No. CIAPOS/2023/406. P.Z. thanks the support received from the Juan de la Cierva fellowship program (JDC2023-050926-I). The XRD experiments were performed at MSPD-BL04 beamline at ALBA Synchrotron with the collaboration of ALBA staff, under proposal number 2024088528. The authors would like to thank Diamond Ligh Source for beamtime (proposal CY40347), and the staff of beamline I15 for assistance with data collection.

# For Table of Contents Use Only

# Comparative high-pressure study on rare-earth entropy fluorite-type oxides

Pablo Botella[a,*], David Vie[b] Leda Kolarek[c], Neha Bura[a], Peijie Zhang[a], Anna Herlihy[d], Dominik Daisenberger[d], Catalin Popescu[f], and Daniel Errandonea[a]

[a]*Departamento de Física Aplicada-ICMUV, MALTA Consolider Team, Universitat de Valencia, 46100 Valencia, Spain*
[b]*Institut de Ciència dels Materials de la Universitat de València, Apartado de Correos 2085, E-46071 València, Spain*
[c]*Faculty of Chemical Engineering and Technology, University of Zagreb, 10000 Zagreb, Croatia*
[d]*Diamond Light Source Ltd., Harwell Science & Innovation Campus, Diamond House, Didcot, OX11 0DE, UK*
[f]*CELLS-ALBA Synchrotron Light Facility, Cerdanyola del Vallés, 08290 Barcelona, Spain*

## Synopsis and TOC

Increasing configurational entropy modulates the high-pressure response of fluorite-type rare-earth oxides. Both $(CePr)O_{2-\delta}$ and $(CePrLa)O_{2-\delta}$ exhibit a compressibility anomaly driven by bond-angle distortion, while La-containing compositions undergo reversible amorphization. Raman data reveal pressure-induced local reordering, establishing entropy as a key variable in oxide robustness under extreme conditions.

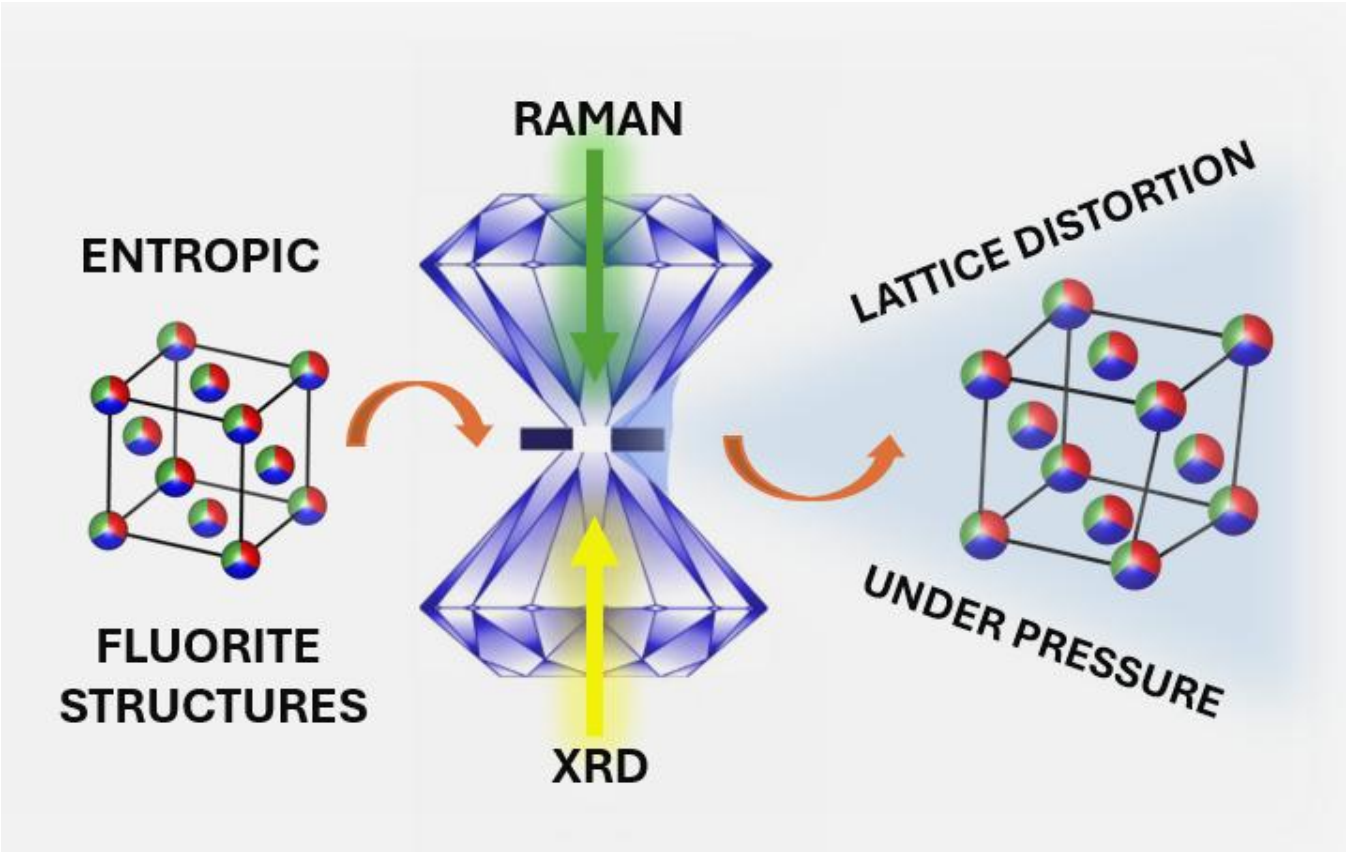